\documentclass[aps,prl,twocolumn,showpacs,10pt,superscriptaddress,preprintnumbers,nofootinbib]{revtex4-1}
\usepackage{epsfig,amssymb,amsmath,psfrag,epstopdf,color}
\usepackage{graphicx}

\allowdisplaybreaks

\def\beq{\begin{equation}}
\def\eeq{\end{equation}}
\def\bsp#1\esp{\begin{split}#1\end{split}}
\newcommand{\be}{\begin{equation}}
\newcommand{\ee}{\end{equation}}
\newcommand{\bea}{\begin{eqnarray}}
\newcommand{\eea}{\end{eqnarray}}

\def\to{\rightarrow}

\def\ksl{\not{\hbox{\kern-2.3pt $k$}}}

\def\Ord{{\cal O}}

\def\alphas{\alpha_s}

\def\spa#1.#2{\left\langle#1\,#2\right\rangle}
\def\spb#1.#2{\left[#1\,#2\right]}
\def\lor#1.#2{\left(#1\,#2\right)}
\def\sand#1.#2.#3{%
\left\langle\smash{#1}{\vphantom1}^{-}\right|{#2}%
\left|\smash{#3}{\vphantom1}^{-}\right\rangle}

\newcommand{\nn}{\nonumber}

\def\sss{\scriptscriptstyle}

\begin{document}

\preprint{MIT-CTP-4758}

\title{Charm-quark production in deep-inelastic neutrino scattering at NNLO in QCD}
\author{Edmond L. Berger}
\email{berger@anl.gov}
\affiliation{High Energy Physics Division, Argonne National Laboratory, Argonne, Illinois 60439, USA}
\author{Jun Gao}
\email{jgao@anl.gov}
\affiliation{High Energy Physics Division, Argonne National Laboratory, Argonne, Illinois 60439, USA}
\author{Chong Sheng Li}
\email{csli@pku.edu.cn}
\affiliation{Department of Physics and State Key Laboratory of Nuclear Physics and
Technology, Peking University, Beijing 100871, China}
\affiliation{Center for High Energy Physics, Peking University, Beijing 100871, China}
\author{Ze Long Liu}
\email{liuzelong@pku.edu.cn}
\affiliation{Department of Physics and State Key Laboratory of Nuclear Physics and
Technology, Peking University, Beijing 100871, China}
\author{Hua~Xing~Zhu}
\email{zhuhx@mit.edu}
\affiliation{Center for Theoretical Physics, Massachusetts
  Institute of Technology, Cambridge, MA 02139, USA}

\begin{abstract}
\noindent
We present a fully differential next-to-next-to-leading order calculation of charm quark
production in charged-current deep-inelastic scattering, with full charm-quark mass 
dependence.  The next-to-next-to-leading order corrections in perturbative quantum 
chromodynamics are found to be comparable in size to the next-to-leading order
corrections in certain kinematic regions.  We compare our predictions with data on 
dimuon production in (anti-)neutrino scattering from a heavy nucleus.  Our results
can be used to improve the extraction of the parton distribution function of a strange quark in the nucleon.
 
\end{abstract}

\maketitle

\noindent \textbf{Introduction.}
Charm-quark ($c$) production in deep-inelastic scattering (DIS) of a neutrino from a heavy-nucleus provides direct access 
to the strange quark ($s$) content of the nucleon.  At lowest order, the relevant partonic process is neutrino interaction with 
a strange quark, $\nu s \rightarrow c X$, mediated by weak vector boson $W$ exchange.  Another source of constraints is 
charm-quark production in association with a $W$ boson at hadron colliders, $g s \rightarrow c W$.  The DIS data determine 
parton distribution functions (PDFs) in the nucleon whose detailed understanding is vital for precise predictions at the Large 
Hadron Collider (LHC).  
The strange quark PDF can play an important role in LHC phenomenology, contributing, for example, to the total PDF uncertainty 
in $W$ or $Z$ boson production~\cite{Nadolsky:2008zw,Kusina:2012vh}, and to systematic uncertainties in 
precise measurements of the $W$ boson mass and weak-mixing angle~\cite{Krasny:2010vd,Bozzi:2011ww,Baak:2013fwa}.
It is estimated that the PDF uncertainty of the strange quark alone could lead to an uncertainty of about 10 MeV
on the $W$ boson mass measurement at the LHC~\cite{wmass}. 
From the theoretical point of view it is important to test whether the strange PDFs are suppressed compared to those of other 
light sea quarks, related to the larger mass of the strange quark, as suggested by various 
models~\cite{Carvalho:1999he,Vogt:2000sk,Chen:2009xy}, and to establish whether there is a difference between the strange and 
anti-strange quark PDFs.   

In this Letter we report a complete calculation at next-to-next- 
to-leading-order (NNLO) in pertubative quantum chromodynamics (QCD) of charm-quark production in DIS of a 
neutrino from a nucleon.  Our calculation is based on a phase-space slicing method and uses fully-differential 
Monte Carlo integration.  It  maintains the exact mass dependence and all kinematic information at the parton level.  
The NNLO corrections can change the cross sections by up to 10\% depending on the kinematic region considered.
Our results show that the next-to-leading order (NLO) predictions underestimate the perturbative uncertainties owing 
to accidental cancellations at that order.  Our calculation is an important ingredient for future global analyses of PDFs at NNLO in QCD, especially for extracting the strange quark PDFs.  The results  can also be 
used to correct for acceptance in experimental analyses.  In this Letter, we show comparisons of our results with 
data from the NuTeV and NOMAD collaborations~\cite{Goncharov:2001qe,Samoylov:2013xoa},
indicating that once the NNLO corrections are included 
slightly higher strangeness PDFs are preferred in the low-$x$ region than those based on a NLO analysis. 
         
Ours is the first complete NNLO calculation of QCD corrections to charm-quark production in weak charged-current 
deep inelastic scattering. In all current analyses which include charm-quark production data in neutrino DIS, the 
hard-scattering cross sections are calculated only at NLO~\cite{Gottschalk:1980rv,Gluck:1997sj,Blumlein:2011zu}
without including an estimation of the remaining higher-order perturbative uncertainties.
Approximate NNLO~\cite{Alekhin:2014sya} results are available for very
large momentum transfer.  However, for neutrino DIS 
experiments~\cite{Goncharov:2001qe,Mason:2006qa,KayisTopaksu:2008aa,Samoylov:2013xoa}, the typical momentum 
transfer is small and the exact charm-quark mass dependence must be
kept~\cite{Alekhin:2014sya,Pire:2015iza}. Recently $\mathcal {O} (\alpha_s^3)$
results~\cite{Behring:2015roa} became available for structure function $xF_3$ at large momentum transfer. 

In the remaining paragraphs we outline the method used in the calculation, present our numerical results showing 
their stability under parameter variation, and then compare with data in the kinematic regions of the experimental acceptance.\\

\noindent \textbf{Method.}
The process of interest is the production of a charm quark in DIS,
 $ \nu_\mu (p_{\nu_\mu}) + N (p_{\sss N}) \to  \mu^-(p_{\mu^-}) + c(p_c) + X(p_{\sss X}) $,
where $X$ represents the final hadronic state excluding the charm quark.
We work in the region where the momentum transfer $Q^2=-q^2= 
-(p_{\nu_\mu} - p_{\mu^-})^2 $ is much larger than the perturbative scale $\Lambda^2_{\rm QCD}$ and 
perturbative QCD can be trusted.  The 
calculation of QCD corrections beyond LO requires proper handling of divergences in loop and phase
space integrals which must be canceled consistently to produce physical results.  Methods based on
subtraction~\cite{hep-ph/9605323,hep-ph/9512328} or phase-space
slicing~\cite{hep-ph/9302225} have been shown to be successful at NLO. 
The NNLO case is less well developed, although several methods
have been proposed~\cite{hep-ph/0409088,hep-ph/0505111,PRLTA.98.222002,1005.0274,1501.07226,1505.04794,1504.02131}.   
For this calculation we employ phase-space slicing at
NNLO~\cite{1210.2808}, which is a generalization of the $q_T$-subtraction concept of Catani and
Grazzini~\cite{PRLTA.98.222002}. Specifically, we use N-jettiness variable of Stewart, Tackmann and Waalewijn~\cite{1004.2489} to
divide the final state at NNLO into resolved and unresolved
regions. Phase-space slicing based on this observable is also
dubbed the N-jettiness subtraction. For recent applications of N-jettiness
subtraction, see Refs.~\cite{Boughezal:2015ded,Campbell:2016yrh}.  We define 
\begin{align}
  \tau =  \frac{2 \, p_X \cdot p_n }{Q^2 + m^2_c} , \quad p_n  =
\Big(\bar{n} \cdot (p_c - q)\Big)    \frac{ n^\mu }{2} 
\end{align}
where $m_c$ denotes the charm quark mass, $n=(1,0,0,1)$ specifies the
direction of the incoming hadron in the center of mass frame,
and $\bar{n}=(1,0,0,-1)$ denotes the opposite direction.
Following Ref.~\cite{1012.4480}, we call $\tau$ 0-jettiness in this
work.  We refer to the region $\tau \ll
1$ as unresolved, while the region $\tau \sim 1$ as resolved. We
discuss the calculation of cross section in these two regions
separately. 

In the unresolved region, $p_{\sss X} \cdot p_n\sim 0$, i.e.,
$p_{\sss X}$ consists of either soft partons, or hard partons collinear to incoming
hadron, or both. Using the machinery of soft-collinear effective
theory~(SCET)~\cite{hep-ph/0005275,hep-ph/0011336,hep-ph/0109045,hep-ph/0202088},
one may show that the cross section in this region
obeys a factorization theorem~\cite{Stewart:2009yx,BGLLZ}:
\begin{align}
  \frac{d  \sigma_{\rm fact.}}{d\tau}  = &\,  \int^1_0 dz\,
\hat{\sigma}_0 (z) \big| C ( Q, m_c, \mu ) \big|^2 \int d\tau_n \,
d\tau_s 
\label{eq:fac}
\\ &
\times \delta( \tau - \tau_n - \tau_s) B_q ( \tau_n, z, \mu) S( \tau_s,n\cdot v,\mu) \nn
\end{align}
where $\hat{\sigma}_0(z)$ is the  LO partonic cross
section for the reaction $s(zp_{\sss N}) + \nu_\mu(p_{\nu_\mu}) \to c(p_c) + \mu^-
(p_{\mu^-})$. $C ( Q, m_c, \mu ) = 1 + \Ord(\alpha_s)$ is the hard Wilson coefficient
obtained from matching QCD to SCET.  It encodes all the short distance
corrections to the reaction.  Collinear radiation and soft
radiation are described by the beam $B_q(\tau_n,
z, \mu)$ and soft functions $S(\tau_s,n \cdot v, \mu)$.   At LO they have
the simple form
\begin{align}
  \label{eq:2}
  B_q(\tau_n, z, \mu) = \delta(\tau_n) f_{s/N}(z,\mu) , 
\quad S(\tau_s,n \cdot v, \mu) = \delta(\tau_s)
\nn
\end{align}
where $f_{s/N}(z,\mu)$ is the PDF.

The factorization formula Eq.~(\ref{eq:fac}) provides a simplified description of the
cross section, fully differential in the leptonic part and heavy quark
part, and correct up to power corrections in $\tau$.  The 0-jettiness
parameter $\tau$ controls the distance away from the strictly unresolved
region, $\tau = 0$. In fixed order perturbation theory, $d
\sigma/d\tau$ diverges as $\alpha_s^k
\ln^{2k-1}\tau/\tau$, as a result of incomplete cancellation of
virtual and real contributions. The strength of SCET approach to
describing the unresolved region is that each individual component in the
factorization formula Eq.~(\ref{eq:fac}) has its own operator definition
and can be computed separately.  

All the ingredients needed in this Letter have been computed through two loops for
different purposes.  Specifically, the
hard Wilson coefficient can be obtained by crossing the corresponding
hard Wilson coefficient calculated for $b \to u W^-$
decay~\cite{JHEPA.0811.065,PHRVA.D78.114028,NUPHA.B811.77,NUPHA.B812.264}.
The two-loop soft function and beam function have been calculated in Refs.~\cite{PHLTA.B633.739,1401.5478}.  
After substituting the two-loop expressions for the individual components into
Eq.~(\ref{eq:fac}), we obtain the desired two-loop expansion of the cross
section in the unresolved region~\cite{BGLLZ}.

In the resolved region, besides the beam jet, there is at least one
additional hard jet with large recoil against the beam.  While we
don't have a factorization formula in this region, the soft and
collinear singularities are relatively simple.  Owing to the presence of
the hard recoil jet, there is at most one parton which can become soft
or collinear.  A singularity of this sort can be handled by the standard
methods used at NLO.  The relevant ingredients are a) one-loop
amplitudes for charm plus one jet production which we take from~\cite{Campbell:2005bb}
and cross check with GoSam~\cite{Cullen:2014yla}, b) the tree-level
amplitudes for charm plus two jet production~\cite{Murayama:1992gi}, and 
c) NLO dipole subtraction terms~\cite{Catani:2002hc} for canceling
infrared singularities between one-loop and tree-level matrix elements.

After introducing an unphysical cutoff parameter $\delta_\tau$, we combine the contributions 
from the two phase space regions,
\begin{align}
  \sigma = \int^{\delta_\tau}_0 \frac{d\sigma_{\rm fact.}}{d\tau}+
  \int^{\tau_{\rm max}}_{\delta_\tau} \frac{d\sigma}{d\tau} + \Ord( \delta_\tau) .
\end{align}
Power corrections in $\delta_\tau$ come from the use of
factorization formula in the unresolved region.  In order to suppress the power 
corrections, a small value of $\delta_\tau$ is required.  On the
other hand, the integrations in both the unresolved and resolved
regions produce large logarithms of the form $\alpha_s^k
\ln^{2k}\delta_\tau$ at N$^k$LO.   The integral over $\tau$ can be done analytically 
in the unresolved region.   In the resolved region, the large logarithms of $\ln \delta_\tau$ 
result from numerical integration near the singular boundary of phase space, resulting in  
potential numerical instability.  A balance has to be reached between suppressing power 
corrections in $\delta_\tau$ and reducing numerical instability.\\

\begin{figure}[!h]
  \begin{center}
  \includegraphics[width=0.45\textwidth]{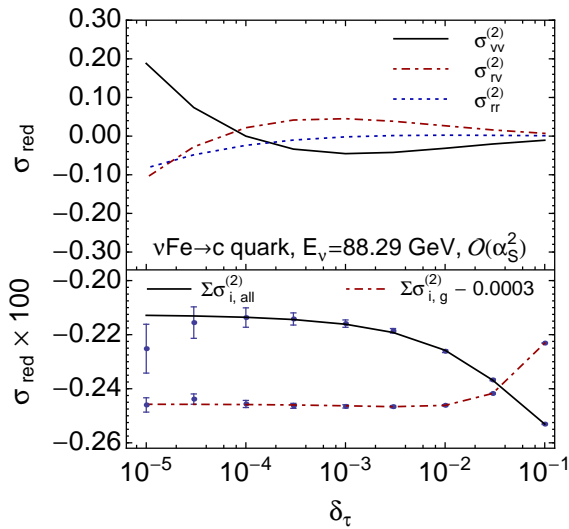}
  \end{center}
  \vspace{-2ex}
  \caption{\label{fig:dscan}
   Dependence of various components of the $\mathcal{O}(\alphas^2)$ reduced cross sections
   on the cutoff parameter for charm-quark production in neutrino DIS from iron.  
   Upper panel: double-virtual (VV), real-virtual (RV), and double-real (RR) contributions to the
   full $\mathcal{O}(\alphas^2)$ corrections; lower panel: the full correction (solid line)  
   and the contribution from the gluon channel shifted by a constant.}
\end{figure}

\noindent \textbf{Numerical results.} 
We first present our numerical results for the total cross section. 
We use CT14 NNLO PDFs~\cite{Dulat:2015mca}
with $N_l=3$ active quark flavors and the associated strong coupling constant.
We use a pole mass $m_c=1.4$ GeV for the charm quark, and CKM matrix elements $|V_{cs}|=0.975$ and
$|V_{cd}|=0.222$~\cite{Beringer:1900zz}.
The renormalization scale is set to $\mu_0=\sqrt{Q^2+m_c^2}$ unless otherwise specified.
In Fig.~\ref{fig:dscan} we plot the NNLO corrections to the reduced cross section~\cite{Mason:2006qa} of charm-quark production in
DIS of neutrino on iron, 
as a function of the phase-space cutoff parameter $\delta_{\tau}$.\footnote{Throughout this paper we do not include 
higher-twist effects, nuclear corrections, electroweak corrections, 
or target-mass corrections. They should be considered when comparing to experimental data and
can be applied separately from the perturbative QCD corrections shown here.}

In the upper panel of Fig.~\ref{fig:dscan} we show three separate contributions to the NNLO corrections: 
the double-virtual part (VV) contributing below cutoff region,
the real-virtual (RV) and double-real (RR) parts contributing to above cutoff
region.  Although the individual contribution vary considerably with
$\delta_\tau$, the total contribution is rather stable and approaches the true NNLO correction when
$\delta_{\tau}$ is small.  The cancellation of the three pieces is about one out of a hundred.
In the lower panel of Fig.~\ref{fig:dscan}, we show the full NNLO correction as well as
its dominant contribution from the gluon channel. Corrections from production initiated by the
strange quark or down quark through off-diagonal CKM matrix elements, and all other
quark flavors, are small comparing to the gluon channel. The error bars indicate the
statistical errors from MC integration and the smooth line is a least-$\chi^2$ fit
of the dependence of the correction on $\delta_{\tau}$.  
As expected the correction is insensitive to the cutoff when $\delta_{\tau}$ is small. 
We find optimal values of 
$\delta_{\tau}$ about $10^{-4}\sim10^{-3}$ for which
the power corrections are negligible while preserving MC integration stability.
According to our fitted results the remaining power corrections there are
estimated to be only a few percents of the NNLO correction itself.  

In neutrino DIS experiments, differential cross sections presented in terms of the Bjorken variable $x$ or the inelasticity $y$.  We 
examined the NLO and NNLO QCD corrections to the differential cross section in $x$ for neutrino scattering on iron, observing that 
the NNLO corrections are comparable to the NLO corrections in the low-$x$ region. When computing the LO, NLO, 
and NNLO cross sections throughout this paper, we consistently use the same NNLO CT14 PDFs~\cite{Dulat:2015mca} in order 
to focus on effects from the matrix elements at the different orders. Decomposing the full corrections 
into different partonic channels, we found that the perturbative convergence is well maintained at NNLO
for gluon or quark channels individually~\cite{BGLLZ}. The NNLO correction to quark channel is much smaller than
at NLO, and the NNLO correction to gluon channel is also below half of the NLO correction. However, at NLO
there is large cancellation between the gluon and quark channels in the small $x$ region.  We regard this cancellation {\em accidental} in that 
it does not arise from basic principles but is a result of several factors.   A similar cancellation has also been observed in the
calculation for $t$-channel single top quark production~\cite{Brucherseifer:2014ama}.  

\begin{figure}[!h]
  \begin{center}
  \includegraphics[width=0.45\textwidth]{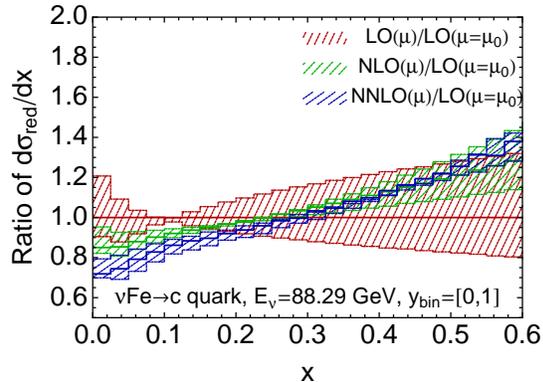}
  \end{center}
  \vspace{-2ex}
  \caption{\label{fig:scale}
   Scale variations at LO, NLO, and NNLO of the differential distribution in $x$ for charm-quark production
   in neutrino DIS from iron, normalized to the LO distribution with the nominal scale choice.
   The solid line shows corresponding central prediction with the nominal scale choice.}
\end{figure}

In Fig.~\ref{fig:scale}, we display the scale variation envelope of the LO, NLO, and NNLO calculations 
for the differential distribution in $x$, normalized to the LO prediction with nominal scale choice.   The 
bands are calculated by varying renormalization and factorization scales together, $\mu_R=\mu_F=\mu$, 
up and down by a factor of two around the nominal scale $\mu_0$, avoiding going below the charm-quark mass.
At low-$x$ the NLO scale variation underestimate the perturbative uncertainties owing to the
accidental cancellations mentioned in the previous paragraph.  The NLO scale variations do not reflect
the size of the cancellations between different partonic channels.  The NNLO
scale variations give a more reliable estimation of the perturbative uncertainties and also
show improvement at high-$x$ compared with the NLO case.  
 
\begin{figure}[!h]
  \begin{center}
  \includegraphics[width=0.45\textwidth]{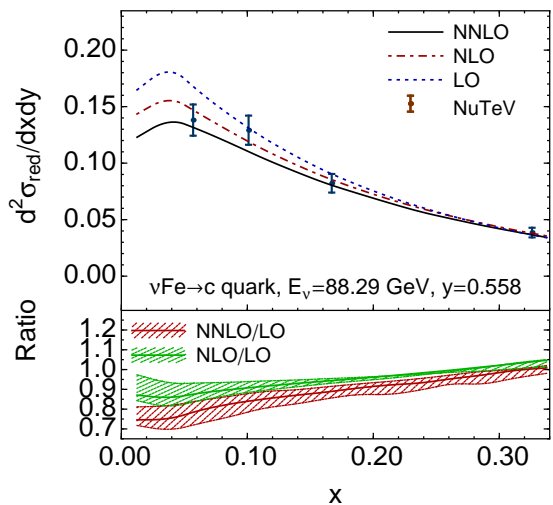}
  \end{center}
  \vspace{-2ex}
  \caption{\label{fig:nutev1}
   Comparison of theoretical predictions to the doubly-differential cross
   sections measured by NuTeV for charm-quark production through neutrino DIS from 
   iron.}
\end{figure}

\noindent \textbf{Comparisons with data.} We turn to an examination of the effects of the NNLO corrections in 
the kinematic regions of two specific neutrino DIS experiments.  The first is the NuTeV collaboration 
measurement of charm (anti-)quark production from (anti)neutrino scattering from 
iron~\cite{Goncharov:2001qe,Mason:2006qa}.  They measure the cross sections for dimuon final states,
where one of the muons arises from the primary interaction vertex and the other one from 
semileptonic decay of the produced charmed hadron.  Kinematic acceptance and the inclusive branching 
ratio to a muon are applied to convert the dimuon cross sections to cross sections of charm (anti-)quark
production at the parton level.  These dimuon data from NuTeV have been included in most
of the NNLO fits of PDFs and have played an important role in constraning strangeness PDFs.
The data are presented as doubly-differential cross sections in $x$ and $y$.  
In Fig.~\ref{fig:nutev1} we show a comparison of theoretical predictions with the 
data for neutrino scattering with $y=0.558$, for several values of $x$.   As expected, 
the NLO calculations generally agree with the data since the same data and the 
same NLO theoretical expressions are used in the CT14 global analyses~\cite{Dulat:2015mca}.  
The NNLO corrections are negative in the region of the data and can be as large as 10\% of the NLO 
predictions, as shown in lower panel of Fig.~\ref{fig:nutev1}.  Based on this comparison, 
we expect that once the NNLO corrections are included in the global analyses fits, the preferred central 
values of strange-quark PDFs will be shifted upward. The shift
represents one of the theoretical systematics that has not yet been taken into account in
any of current global analyses.

\begin{figure}[!h]
  \begin{center}
  \includegraphics[width=0.45\textwidth]{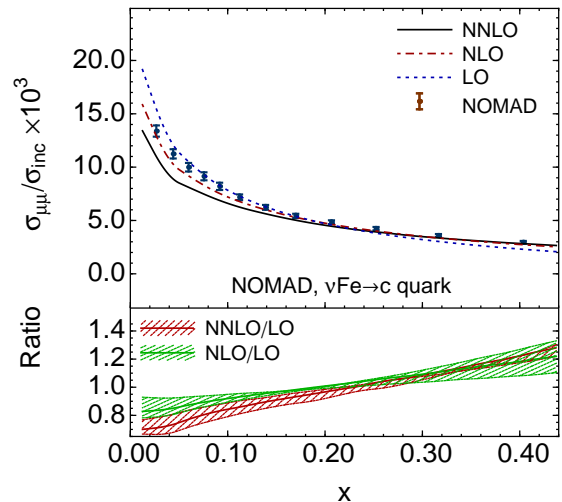}
  \end{center}
  \vspace{-2ex}
  \caption{\label{fig:nomad}
   Comparison of theoretical predictions for ratios of the dimuon cross section to
   the inclusive charged-current cross section measured by NOMAD for neutrino DIS from 
   iron.}
\end{figure}

The second set of data is from the NOMAD collaboration measurement of neutrino scattering from 
iron~\cite{Samoylov:2013xoa}.  They present ratios of dimuon cross sections to inclusive charged-current cross
sections $\mathcal{R}_{\mu\mu}\equiv\sigma_{\mu\mu}/\sigma_{inc}$ instead of converting the dimuon cross sections
back to charm-quark production.   The measurement is done with a neutrino beam of continuous energy peaked 
around 20 GeV.  A $Q^2$ cut of 1 ${\rm GeV}^2$ has been applied.
In Fig.~\ref{fig:nomad} we show our comparisons of predictions to data as a function
of $x$.  Here we consistently use the NNLO results for $\sigma_{inc}$ in the 
denominator of the ratio, obtained from the program
OPENQCDRAD~\cite{openqcdrad,Blumlein:2014fqa}.
By LO, NLO, and NNLO in the figure we refer to our calculations of the dimuon cross sections in the
numerator of the ratio.  The NLO calculations generally agree with data although these data are not 
included in the CT14 global analyses. The NNLO corrections are negative and can reach 
about 10\% of the LO cross sections in the low-$x$ region covered by the data.
At high $x$ the NNLO corrections are only a few percent and become
positive. The NNLO corrections in  Fig.~\ref{fig:nomad} are
generally larger than the experimental errors.  Thus, they can
be very important for extracting strange-quark PDFs in
analyses with NOMAD data included. We also plot
the scale variation bands in lower panel of Fig.~\ref{fig:nomad}.
The trends are similar to ones in Fig.~\ref{fig:scale}. The NLO
predictions underestimate the perturbative uncertainty. It can
still reach $\pm 5$\% at NNLO in the low-$x$ region and can be
reduced once even higher order corrections are included.
     
\noindent \textbf{Summary.} We present the first complete calculation of NNLO QCD corrections to
charm-quark production in weak charged-current deep inelastic scattering.
The calculation is fully differential based on a generalization of phase-space slicing to NNLO in QCD.  
The NNLO corrections can change the cross sections by up to 10\% depending on the kinematic region considered.
In the kinematic regions of the NuTeV and NOMAD collaborations our results indicate that once the NNLO
corrections are included, the data prefer slightly larger strangeness PDFs in the low-$x$ region than
those based on NLO predictions.  A definitive result awaits a global analysis with the NNLO corrections included, 
left for a future study.\\

\begin{acknowledgments}
Work at ANL is supported in part by the U.S. Department of Energy
under Contract No. DE-AC02-06CH11357. 
HXZ was supported by the Office of Nuclear Physics of the U.S.DOE under Contract No. DE-SC0011090.
This work was also supported in part by the National Nature Science Foundation of China, under
Grants No. 11375013 and No. 11135003.
We thank Pavel Nadolsky for valuable comments and Southern Methodist University for
the use of the High Performance Computing facility ManeFrame. \\
\end{acknowledgments}

\end{document}